\documentclass[aps,reprint]{revtex4-1}
\usepackage{graphicx}% Include figure files
\usepackage{dcolumn}% Align table columns on decimal point
\usepackage{bm}% bold math
\usepackage{amsmath}
\usepackage[dvipsnames]{xcolor}
\draft % marks overfull lines with a black rule on the right

\begin{document}

%\preprint{}

\title{Measuring magnetic flux suppression in high-power laser-plasma interactions}

\author{P. T. Campbell}
\email{campbpt@umich.edu}
\affiliation{G\'{e}rard Mourou Center for Ultrafast Optical Science, University of Michigan, 2200 Bonisteel Boulevard, Ann Arbor, Michigan 48109, USA}
\author{C. A. Walsh}
\affiliation{Lawrence Livermore National Laboratory, Livermore, California 94550, USA}
\author{B. K. Russell}
\affiliation{G\'{e}rard Mourou Center for Ultrafast Optical Science, University of Michigan, 2200 Bonisteel Boulevard, Ann Arbor, Michigan 48109, USA}
\author{J. P. Chittenden}
\author{A. Crilly}
\affiliation{Blackett Laboratory, Imperial College, London SW7 2AZ, United Kingdom}
\author{G. Fiksel}
\affiliation{G\'{e}rard Mourou Center for Ultrafast Optical Science, University of Michigan, 2200 Bonisteel Boulevard, Ann Arbor, Michigan 48109, USA}
\author{L. Gao}
\affiliation{Princeton Plasma Physics Laboratory, Princeton University, Princeton, New Jersey 08543, USA}
\author{I. V. Igumenshchev}
\author{P. M. Nilson}
\affiliation{Laboratory for Laser Energetics, 250 East River Road, Rochester, New York 14623, USA}
\author{A. G. R. Thomas}
\author{K. Krushelnick}
\author{L. Willingale}
\affiliation{G\'{e}rard Mourou Center for Ultrafast Optical Science, University of Michigan, 2200 Bonisteel Boulevard, Ann Arbor, Michigan 48109, USA}

\date{\today}

\begin{abstract}

Biermann battery magnetic field generation driven by high power laser-solid interactions is explored in experiments performed with the OMEGA EP laser system.
Proton deflectometry captures changes to the strength, spatial profile, and temporal dynamics of the self-generated magnetic fields  as the target material or laser intensity is varied. Measurements of the magnetic flux during the interaction are used to help validate extended magnetohydrodynamic (MHD) simulations.
Results suggest that kinetic effects cause suppression of the Biermann battery mechanism in laser-plasma interactions relevant to both direct and indirect-drive inertial confinement fusion. Experiments also find that more magnetic flux is generated as the target atomic number is increased, which is counter to a standard MHD understanding.

\end{abstract}

\maketitle

\section{\label{sec:intro}Introduction}
High power laser-solid interactions can create high energy density (HED) plasma conditions relevant to inertial confinement fusion (ICF) and laboratory astrophysics research \cite{Drake_HEDP,Remmington_RevModPhys_2006}.
In laser-produced plasmas, strong magnetic fields can be spontaneously generated by a number of mechanisms \cite{Haines_CJP_1986}, though the primary source is the Biermann battery effect caused by nonparallel temperature and density gradients ($
\partial B/\partial t \propto \nabla T_e \times \nabla n_e$) \cite{Biermann_PR_1951,Stamper_PRL_1971,Stamper_PRL_1975}.
A detailed understanding of self-generated magnetic fields is critical to laser-fusion research because strong fields can influence thermal energy transport \cite{Braginskii_1965}, and potentially impact the evolution of hydrodynamic instabilities \cite{Evans_PPCF_1986, Manuel_PRL_2012,  Gao_PRL_2012_RT_side-on, Gao_PRL_2013_RT_face-on,Hill_PRE_2018}.
Laser-driven magnetic fields also enable laboratory investigations of magnetized astrophysical phenomena, especially magnetic reconnection \cite{Nilson_PRL_2006,Li_PRL_2007,Willingale_POP_2010,Zhong_NatPhys_2010,Fox_PRL_2011,Dong_PRL_2012}.

For both ICF and laboratory astrophysics research, numerical modeling is an essential predictive tool, and the extended-magnetohydrodynamics (extended-MHD) framework has been developed to describe transport of energy and magnetic fields in HED plasmas \cite{Walsh_POP_2020}.
Recent simulation work has shown that extended-MHD effects such as Nernst and Righi-Leduc can modify plasma properties in indirect-drive ICF hohlraums \cite{Farmer_POP_2017}, direct-drive ICF ablation fronts \cite{Hill_PRE_2018}, and at the edge of compressed fusion fuel \cite{Walsh_PRL_2017}.
Accurate extended-MHD modeling is crucial to development and interpretation of advanced ICF experiments with pre-magnetized fuel \cite{Davies_POP_2015_premag,McBride_POP_2015,Joglekar_PRE_2016,Walsh_POP_2019_premag}.
In addition, simulations anticipate that extended-MHD effects can dictate reconnection rates in laser-driven magnetic reconnection experiments \cite{Joglekar_PRL_2014,Tubman_NatComm_2021}.

Though relatively simple in the broader context of HED experiments, a single laser spot interacting with foil targets can provide a powerful platform for validating extended-MHD modeling.
Proton deflectometry enables high spatial and temporal resolution measurements of magnetic field generation and dynamics in the laser-produced plasma \cite{Borghesi_PPCF_2001,Li_PRL_2006}.
Using moderate laser intensities,  $I_L = 10^{14} - 10^{15}$ Wcm$^{-2}$, recent experiments have demonstrated that simulations of laser-foil interactions must incorporate key physical processes such as Biermann battery field generation, cross-field Righi-Leduc heat flow, and Nernst advection \cite{Willingale_PRL_2010,Lancia_PRL_2014,Gao_PRL_2015}.
The Nernst effect moves fields down temperature gradients ($\textrm{v}_\textrm{N} \propto -T_e^{3/2}\nabla T_e$) \cite{Nishiguchi_PRL_1984,Kho_PRL_1985,Ridgers_PRL_2008}.
In laser-produced plasmas, the Nernst effect can convect magnetic fields with the heat flow toward the ablation region, counter to the bulk plasma flow into the corona \cite{Haines_PPCF_1986}.
Measurements of the magnetic field dynamics can be used to diagnose temperature and density gradients in the plasma, and interplay between energy transport and field generation.

By varying the target material, the effect of atomic or radiation physics on transport and field dynamics can be explored.
Recent work using proton deflectometry captured distinct regions of magnetic field generation around radiation-driven double ablation fronts (DAF) in mid-Z targets \cite{Campbell_PRL_2020}.
Incorporating radiation transport into extended-MHD simulations reproduced the DAF formation and concentric double field features.

In that work, it was found that extended-MHD simulations overestimated the magnetic field strength. 
It is anticipated that non-local effects not captured by the extended-MHD framework can suppress the rate of Biermann battery field generation in regions where the electron mean free path ($\lambda_{ei}$) approaches (or exceeds) the local temperature gradient length scale ($l_T = (T_e/\nabla T_e) $) \cite{Sherlock_PRL_2020,Hill_PRE_2018}.
Using empirical fits to kinetic simulations, Sherlock and Bissell \cite{Sherlock_PRL_2020} developed scalings for the suppression of classical Biermann battery generation rates as a function of the ratio $\lambda_{ei}/l_T$.
However, the results have not yet been compared to experiments until now.

In this paper, high resolution proton deflectometry measurements quantify target material effects on magnetic fields generated during high power laser-solid interactions.
\textbf{}Experimental observations of magnetic flux are used to help validate extended-MHD simulations that include the new scalings for non-local suppression of Biermann battery field generation, as well as radiation transport.

\section{Methods}
\subsection{Experiments}

In this work, data is drawn from two separate experimental campaigns performed with the OMEGA EP laser system at the University of Rochester's Laboratory for Laser Energetics.
Magnetic field generation was driven by either one \cite{Campbell_PRL_2020} or two overlapped \cite{Gao_PRL_2015} UV laser pulses ($\lambda_L = 351$~nm) interacting with thin foil targets.
In the single pulse case, a beam with 1.25~kJ of energy and 1~ns square temporal profile was focused using a distributed phase plate on to an 819~$\mu$m diameter ($d_{95}$) super-Gaussian spot with a $\sim 30^{\circ}$ angle of incidence to produce an intensity of $2.2\times10^{14}$~Wcm$^{-2}$.
The foil target material was varied between 50~$\mu$m thick plastic (CH), 25~$\mu$m copper, 25~$\mu$m aluminum, or 50~$\mu$m aluminum coated with either 1~$\mu$m of copper (Cu+Al) or gold (Au+Al).

\begin{figure*}[t]
 	\centering
 	\includegraphics[width = \textwidth]{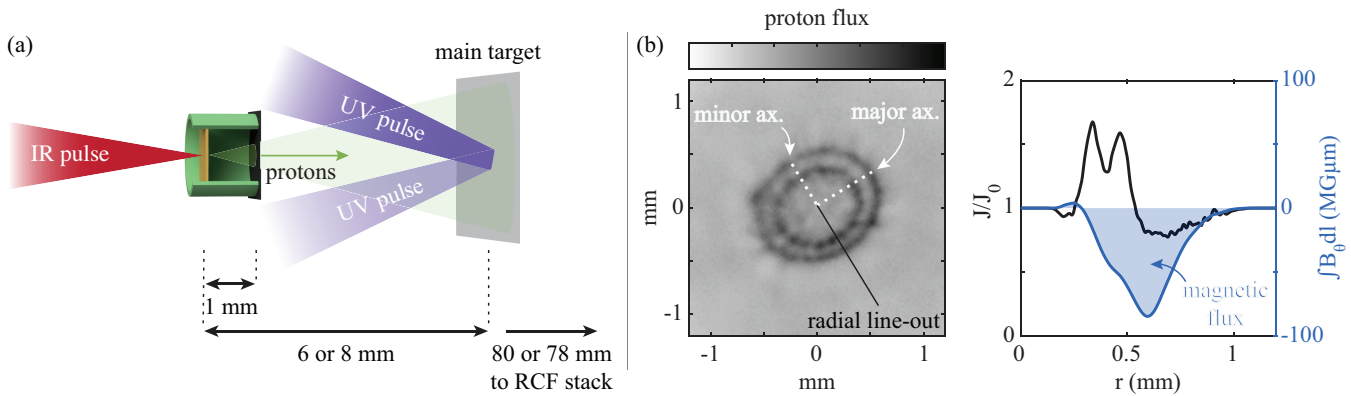}
 	\caption{(a) Schematic of the OMEGA EP experimental setup. 
 	(b) An illustration of the proton image analysis method. Due to the laser angle-of-incidence, the field features are mildly elliptical. Line-integrated magnetic field profiles are reconstructed from normalized radial line-outs (J/J$_0$) taken along minor axis. Integrating the field profile in the radial direction yields a measurement of the magnetic flux.
 }
 	\label{fig:bfield_setup}
\end{figure*}

In the two beam experiment, each pulse contained 2~kJ in a 2.5~ns square temporal profile.
The pulses were overlapped onto an 734~$\mu$m diameter spot with a $23^{\circ}$ angle of incidence to produce an combined intensity of $4.4\times10^{14}$~Wcm$^{-2}$.
For this series of shots, the targets were limited to 50~$\mu$m thick plastic foils.

A schematic of the experimental geometry is shown in Figure~\ref{fig:bfield_setup}(a).
Self-generated magnetic fields were imaged by protons in a point-projection geometry with magnifications ranging from $\sim$10--14.
In both experiments, the high energy proton probe was produced via the target normal sheath acceleration (TNSA) mechanism \cite{Maksimchuk_PRL_2000,*Snavely_PRL_2000,*Clark_PRL_2000,*Hatchett_POP_2000}.
A short infrared (IR) laser pulse (300~J, 0.7--1~ps) was focused to intensities exceeding $10^{19}$ Wcm$^{-2}$ onto 20--50~$\mu$m thick copper foils, accelerating protons with a quasi-Maxwellian energy spectrum and maximum energies around 60 MeV.
To protect the proton source from coronal plasma or x-ray preheat emitted from the main interaction, the foil was mounted within a plastic tube capped by a 5~$\mu$m thick tantalum shield.
Because of the laminar propagation, small virtual source size ($\sim 10$~$\mu$m), and short emission duration ($\sim 1$~ps), proton beams from a TNSA source enable high spatial and temporal resolution imaging of self-generated electric and magnetic fields \cite{Borghesi_PRL_2004}.
The relative timing between the main interaction and the proton probe could be adjusted with $\pm20$~ps error to measure the temporal evolution of the field features.

The proton beams were detected with filtered stacks of radiochromic film (RCF).
Due to a combination of the Bragg peak in proton energy deposition and a thin sensitive region, each RCF layer detects a narrow energy slice of the accelerated spectrum ($\Delta E/E \leq 1\%$).
Deflections from fields generated in the main interaction will result in proton fluence modulations on the film.
Quantitative measurements of path-integrated field strengths can be retrieved from the relative distribution of protons compared to the undisturbed beam profile \cite{Borghesi_PPCF_2001,Li_PRL_2006,Kugland_RSI_2012}.

A 1D polar-coordinates field reconstruction technique was developed to extract quantitative path-integrated magnetic field information from radial line-outs through the proton images.
A detailed description of the reconstruction method can be found in the Supplemental Material of Ref.\ \onlinecite{Campbell_PRL_2020}.
The proton image analysis method is illustrated in Figure~\ref{fig:bfield_setup}(b). 
In this probing geometry, protons are primarily sensitive to azimuthal magnetic fields generated in the laser-produced plasma \cite{Li_PRL_2006}.
Due to the laser angle of incidence, the observed features are elliptical.
Typically, radial line-outs were taken along the minor axis for best comparison to the 2D (r-z) simulations described below.

A key challenge and source of error in the field reconstruction method is accurate determination of the undisturbed proton profile, J$_0$.
Shot-to-shot fluctuations in the undeflected beam profile and fluence means that direct use of reference data taken from other shots is ineffective.
Instead, the low spatial frequency undisturbed profiles were inferred from the line-out signal (J) using Fourier filtering.
(Note: in this work, the Fourier filter is applied to a line-out containing the full diameter of the proton image feature)
After Gaussian low-pass filtering, the signal level is adjusted such that total proton flux is conserved ($\sum$J $= \sum$ J$_0$).
The reconstructed field profiles are constrained by assuming that the field strength should drop to zero near the center of the focal region and the outer edge (far from the interaction).
This is accomplished by using a super-Gaussian mask to blend the filtered signal with the original line-out such that $J/J_0 = 1$ as $r < r_{min}$ and $r > r_{max}$.

For each line-out, a scan of Gaussian low-pass filter widths, $r_{min}$, and $r_{max}$ values are tested. 
Filter widths range from 0.4 to 3~mm, with 0.05~mm spacing (from approximately the diameter of ring features to the full width of line-out).
After visually selecting starting points, $r_{min}$ and $r_{max}$ are varied over $\pm 0.05$~mm with spacing of 0.025~mm.
The combination of these parameters produces a grid of possible undisturbed profiles (J$_0$).
For each inferred J$_0$, a path-integrated field profile is reconstructed and a subset of solutions is selected based on the criteria that $\int \!\! B_{\theta} \textrm{d}z  \rightarrow 0$ for $r> r_{max}$.
An example result of analysis approach is shown in Figure~\ref{fig:bfield_setup}(b).
The mean normalized fluence (J$/$J$_0$) and mean reconstructed path integrated magnetic field are plotted together.
An observable that will be compared to the simulations is the magnetic flux, calculated by integrating in the radial direction.

A number of approaches are used to characterize the uncertainty of the field reconstruction: taking line-outs at different angles, using a larger range of values for $r_{max}$, analyzing of successive layers of RCF (the relative proton time-of-flight differences are small compared to the interaction time scale), or artificially suppressing RCF signal to approximate uncertainties in the RCF sensitivity.
The relative influence of the RCF signal is low because the reconstruction is calculated using the normalized fluence (J/J$_0$).
Overall, uncertainty in inferring J$_0$ and analysis of successive RCF layers leads to an error of $\sim25\%$ in the path-integrated field strength and magnetic flux.
In addition, the accuracy of the measurement is potentially impacted by blurring due to small-angle proton scattering, especially in the higher Z targets, and by enhanced proton stopping in laser-heated regions of the targets.

\subsection{Extended-Magnetohydrodynamics simulations}

Experimental results were compared to extended-MHD simulations performed using the Gorgon code  \cite{Chittenden_PPCF_2004_GORGON, Ciardi_POP_2007_GORGON, Walsh_PRL_2017} to help validate modeling of magnetic field generation. The evolution of magnetic field in the code is \cite{Walsh_POP_2020}:

\begin{align}
\begin{split}
\frac{\partial \underline{B}}{\partial t} = & - \nabla \times \frac{\alpha_{\parallel}}{\mu_0 e^2 n_e ^2} \nabla \times \underline{B} + \nabla \times (\underline{v}_B \times \underline{B} ) \\
&+ \nabla \times \Bigg( \frac{\nabla P_e}{e n_e} - \frac{\beta_{\parallel} \nabla T_e}{e}\Bigg) \label{eq:mag_trans_new}
\end{split}
\end{align}

Where the first term on the right represents resistive diffusion with coefficient $\alpha_{\parallel}$ and the second term is advection of the magnetic field at velocity $\underline{v}_B$:

\begin{equation}
\label{eq:mag_trans_new_velocity}	\underline{v}_B = \underline{v} - \gamma_{\bot} \nabla T_e - \gamma_{\wedge}(\underline{\hat{b}} \times \nabla T_e)   
\end{equation}

i.e. the magnetic field is advected by bulk plasma motion, Nernst and cross-gradient-Nernst advection. $\gamma_{\bot}$ and $\gamma_{\wedge}$ are magnetic transport coefficients with a similar form to the associated thermal conductivities \cite{Walsh_submitted_nuclearfusion}. $\underline{\hat{b}}$ is the magnetic field unit vector.

The final two terms in equation \ref{eq:mag_trans_new} are the only sources of magnetic flux in the simulations. $\nabla \times (\nabla P_e /en_e)$ is the Biermann battery term and is the dominant source of magnetic flux in the simulations. $\nabla \times \beta_{\parallel} \nabla T_e /e$ represents magnetic flux generated by ionization gradients in the plasma \cite{Sadler_PhilTrans_2020}. Previous Gorgon simulations of foils did not include this term \cite{Campbell_PRL_2020}, although it is found here to be of only secondary importance. 

Magnetic field generation in a laser absorption region has long been anticipated to be suppressed by kinetic processes. 
This is supported by both experiments \cite{Igumenshchev_POP_2014,Gao_PRL_2015} and Vlasov-Fokker-Planck (VFP) simulations \cite{Sherlock_PRL_2020,Hill_PRE_2018}.
Recent VFP simulations of laser absorption regions proposed an empirical fit to the VFP data \cite{Sherlock_PRL_2020}:

\begin{equation}
    \frac{\partial B}{\partial t} \approx 0.083 \bigg(
    \frac{l_T}{\lambda_{ei}}\bigg)^{0.453} \bigg(\frac{\partial 
    B}{\partial t} \bigg)_{\textrm{classical}} \label{eq:sherlock_sup}
\end{equation}

Where $l_T$ is the temperature length-scale and $\lambda_{ei}$ is the mean free path of a thermal electron.  
Equation \ref{eq:sherlock_sup} was proposed by computational work and yet to be compared with experimental results; this is one of the primary goals of this paper. 
Equation \ref{eq:sherlock_sup} has been implemented into Gorgon as an option, and is referred throughout this paper as `Biermann suppression'. 
The ratio $(l_T/\lambda_{ei})$ is calculated at every point across the simulation domain, and the suppression factor is limited to maximum value of 1 (where $l_T/\lambda_{ei} > 243$) so that the generation approaches the classical rate.
Of course, there are limitations to such an approach. This suppression behaviour was observed using a prescribed laser intensity and varying the transverse length-scale of the laser non-uniformity. 
To obtain a complete picture, the VFP simulations would need to vary all initial conditions, which is not practical.
Equation \ref{eq:sherlock_sup} also assumes that the plasma is in a kinetic steady state.
In practice it takes time for the electron distribution function to be perturbed away from Maxwellian.  

Suppression of the Nernst term by kinetic processes has also been proposed \cite{Sherlock_PRL_2020} in a simple form similar to equation \ref{eq:sherlock_sup}. This has also been included as an option in the Gorgon simulations, but does not qualitatively change the results shown here. For simplicity, Nernst suppression has been omitted from the presented work. In this way the connection between the heat-flow and magnetic transport can be maintained \cite{Walsh_POP_2020,Sadler_PRL_2021}. The authors suppose that if the Nernst term is being suppressed using a simple mean-free path argument, then the heat-flow should also be treated similarly.

The heat-flow in Gorgon is fully anisotropic and includes the Righi-Leduc term. The transport coefficients have been updated \cite{Sadler_PRL_2021} and now exhibit physical behaviour at low magnetizations, unlike the Epperlein \& Haines coefficients \cite{Epperlein_and_Haines}. For the configuration simulated here, this mainly lowers the importance of the cross-gradient-Nernst term ($\gamma_{\wedge}$ in equation \ref{eq:mag_trans_new_velocity}).

Simulations in this paper are exclusively two-dimensional, invoking cylindrical symmetry ($r$-$z$). The laser propagates along $z$ with an assumption that the laser is symmetric in $\theta$. Laser propagation uses a simple ray-trace scheme with inverse-bremsstrahlung heating of the electron population. 
Gorgon uses the Frankfurt equation of state (FEoS) with a Thomas-Fermi ionization model, and implements multi-group non-diffusive radiation transport using a P$_{1/3}$ automatic flux-limiting method. 
For CH, 54 radiation energy groups are used, while 300 groups are used for copper.

\begin{figure}
 	\centering
 	\includegraphics[]{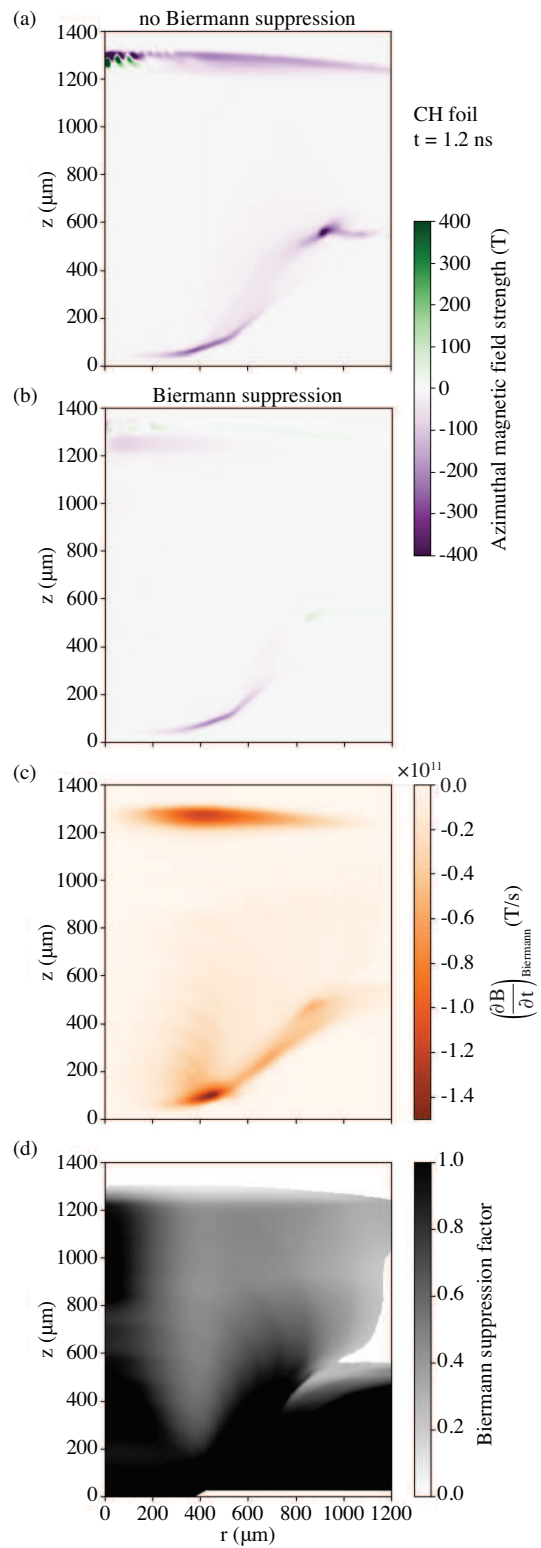}
 	\caption{Simulation results with a CH target and the higher laser intensity at t = 1.2~ns. (a-b) show the magnetic fields without and with Biermann battery suppression included, (c) the instantaneous classical Biermann battery generation rate, and (d) the Biermann suppression factor (lower numbers signify more suppression).
 }
 	\label{fig:sims}
\end{figure}

Figure \ref{fig:sims} demonstrates the impact of Biermann battery suppression in the Gorgon simulations. All of the images are at 1.2~ns for CH using the higher laser intensity \cite{Gao_PRL_2015}. The first two images show the magnetic field structure without and with Biermann battery suppression included in the code. The third image then shows the classical Biermann battery generation rate at this instant in time. The final image shows the suppression factor (equation \ref{eq:sherlock_sup}). The kinetic suppression is particularly important deep into the corona, where the plasma is hot and low in density (resulting in long electron mean-free paths).

For the case without Biermann suppression included, an instability that generates magnetic field is observed in the corona, giving oscillating polarities near the laser axis. These oscillations do not contribute to the overall quantification of magnetic flux, as they cancel out when summed. The instability is likely due to the interplay between Biermann battery and anisotropic thermal conduction, which is called the field-generating thermal instability \cite{Tidman_POF_1974}. It could also be from the magneto-thermal instability \cite{Bissell_PRL_2010}, which results from the interplay between Nernst and Righi-Leduc. Nonetheless, kinetic suppression of the Biermann battery process is found to stabilize the instability, which has been noted in full VFP simulations \cite{Sherlock_PRL_2020}.

Subsequent sections will compare these simulation results to experiments using different laser intensities and foil materials. The comparison suggests that kinetic suppression of Biermann battery generation is indeed occurring in experiments.

\section{Results and Discussion}

\begin{figure}
    \centering
    \includegraphics[]{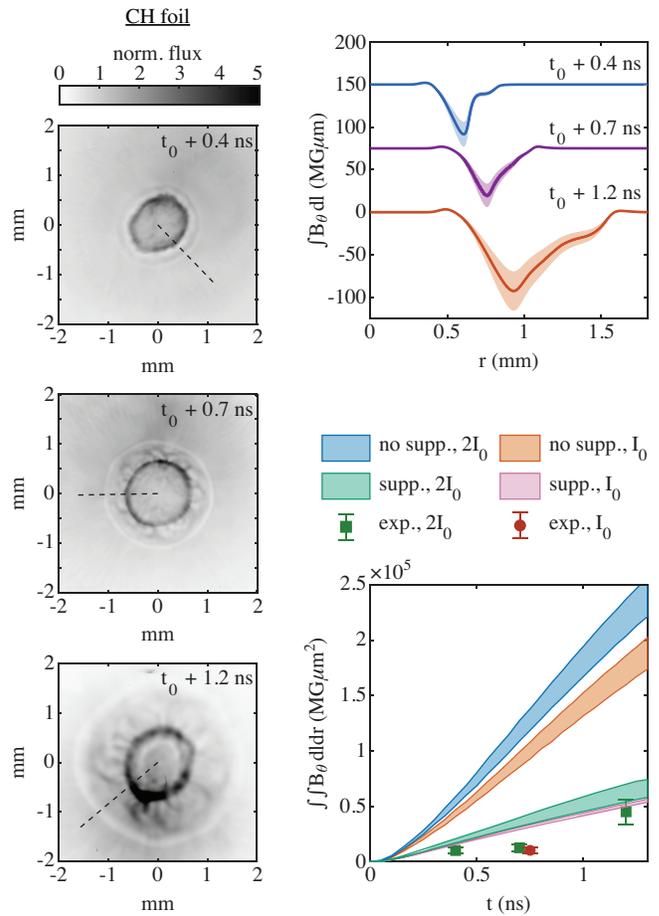}
    \caption{Comparison of experimental and simulation results for CH foils.
    The left column shows proton deflectometry images of fields driven by the higher, overlapped laser intensity (2I$_0$) taken at 0.4~ns, 0.7~ns and 1.2~ns.
    Line-out locations are indicated by dashed lines.
    Reconstructed magnetic field profiles are plotted in the top right panel (offset vertically for clarity).
    In the bottom right panel, the magnetic flux predictions from simulations both without and with Biermann suppression for each laser intensity are compared to experimental measurements. 
    Upper and lower bounds on the simulation results are produced by tuning the laser energy to approximate the influence of energy coupling efficiency.}
    \label{fig:CH_flux}
\end{figure}

Experimental and simulation results for CH foil targets are compared in Figure~\ref{fig:CH_flux}.
In the left column, proton deflectometry images show the evolution of magnetic field structures using the higher laser intensity (2I$_0$, overlapped pulses).
The primary features are concentric light and dark rings of proton fluence due to deflections from an azimuthal magnetic field.
As the plasma evolves, there is evidence in the proton deflectometry images of instability formation with an accompanying electromagnetic field.
Such instability features are 3D in nature, and cannot be captured by the 2D simulations.
For this comparison, line-out locations --- shown with dashed lines --- were chosen to avoid strong modulations caused by the instability (rather than along the minor axis as described in the Methods section above).
The proton energies for these images are 20~MeV for t$_0$+0.4~ns and t$_0$+0.7~ns, and 9~MeV for t$_0$+1.2~ns, where $t_0$ denotes the beginning of UV laser irradiation.

Corresponding reconstructed magnetic field profiles are plotted in the top panel of the right column of Figure~\ref{fig:CH_flux}, with shaded regions showing 25$\%$ uncertainty range.
For clarity, the field profiles for each probing time are offset vertically.
Over the 1.2~ns evolution, the field grows to peak path-integrated strengths near 100~MG$\mu$m.
In qualitative agreement with Ref.\ \onlinecite{Gao_PRL_2015}, the reconstructed profiles indicate that the outer edge of the field expands near the sound speed, 0.8--1$\times 10^6$~m/s, while the largest fields are more closely bound to the focal region and expand more slowly, 0.3--0.5$\times 10^6$~m/s.

The bottom panel of Figure~\ref{fig:CH_flux} compares the evolution of the azimuthal magnetic flux from the experiment and extended-MHD simulations for both laser intensities.
As in the experiment, the higher intensity simulations (2I$_0$) also used a reduced focal spot radius.
Experimental data points are plotted along with shaded regions which show the simulated flux either without or with Biermann suppression included (note: on this y-axis scale, experimental error bars can fall within the plot markers).
The width of the shaded regions illustrate the influence of tuning the laser energy to approximate uncertainty in coupling efficiency for the inverse-bremsstrahlung heating, with upper and lower bounds corresponding to $\sim$90$\%$ and $\sim$70$\%$ coupling.
For both laser intensities, simulations without Biermann suppression greatly overestimate the magnetic flux ($>5\times$).

Agreement is significantly improved by including Biermann suppression, indicating that this effect is likely influencing the field dynamics.
In the simulations, the suppression results in a 3--4$\times$ reduction in the predicted magnetic flux.
For the lower intensity, the inclusion of Biermann suppression also weakens the influence of laser coupling efficiency in the simulation, reducing the percent spread in flux from 15$\%$ to 6$\%$.
The mean difference predicted flux for the two intensities is also reduced by the Biermann suppression, from 25$\%$ to 20$\%$. 
This suggests that achieving higher temperatures with more laser energy is partially balanced by an increase in non-local effects.
Comparing data at t$_0$+0.7~ns for 2I$_0$ and t$_0$+0.75~ns for I$_0$, the experimental difference is 23$\%$, though measurement uncertainty leads to 100$\%$ error in quantifying the relative change.

\begin{figure}
    \centering
    \includegraphics[]{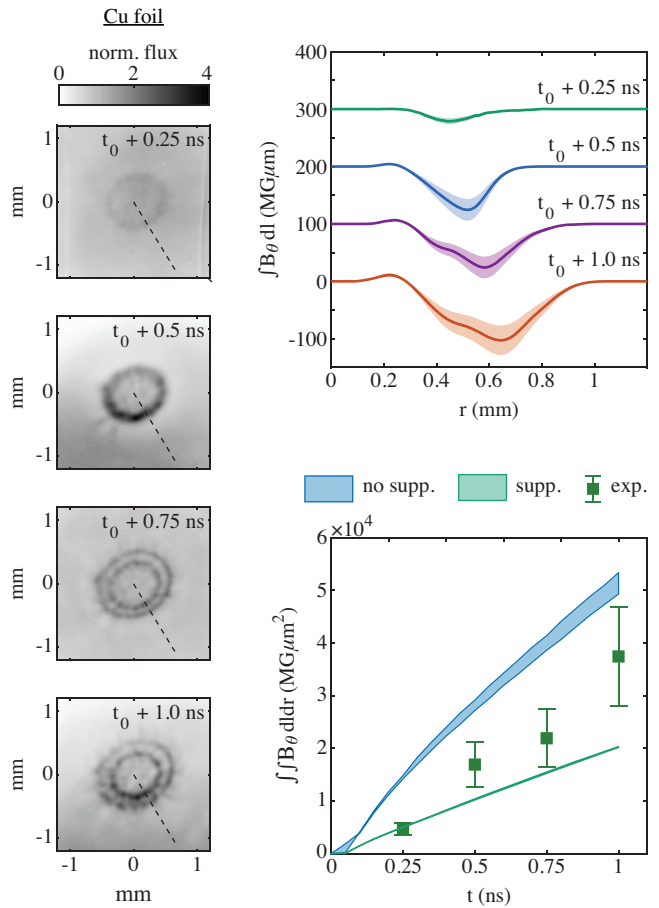}
    \caption{Comparison of experimental and simulation results for Cu foils.
    The left column shows proton images at 0.25~ns, 0.5~ns, 0.75~ns and 1.0~ns.
    Line-out locations are indicated by dashed lines.
    The target for t$_0$+0.25~ns was a 25~$\mu$m-thick Cu foil, and the other probing times use a Cu+Al layered target.
    Reconstructed magnetic field profiles are plotted in the top right panel.
    In the bottom right panel, the magnetic flux predictions from simulations both without and with Biermann suppression are compared to experimental measurements.
    }
    \label{fig:Cu_flux}
\end{figure}

Figure~\ref{fig:Cu_flux} summarizes the results for Cu foil targets with the lower intensity.
The left column shows experimental proton images taken at times ranging from t$_0$+0.25~ns to t$_0$+1.0~ns.
Here, line-outs were taken along the minor axis at each probing time.
The corresponding proton energies are 33.6~MeV for t$_0$+0.25~ns, 22.6~MeV for t$_0$+0.5~ns, and 32.8~MeV for t$_0$+0.75~ns and t$_0$+1.0~ns.
The target for t$_0$+0.25~ns was a 25~$\mu$m-thick Cu foil, while the later measurements were made using the Cu+Al layered target in order to improve the imaging resolution by reducing the effect of scattering.
Previous side-by-side comparisons of image features from pure Cu and layered Cu+Al presented in Ref.\ \onlinecite{Campbell_PRL_2020} indicate that the field dynamics are dominated by the Cu layer. 

Reconstructed path-integrated magnetic field profiles for each probing time are plotted together in the top right panel of Figure~\ref{fig:Cu_flux}.
The field profiles show the expansion of the field features and the emergence of double-peaked structure at t$_0$+0.75~ns and t$_0$+1.0~ns.
This produces the pattern of two concentric rings of proton accumulation observed in the images.
As discussed in Ref.\ \onlinecite{Campbell_PRL_2020}, the two field features are evidence of Biermann battery generation around radiation-driven double ablation front (DAF) structures.

The bottom right panel compares experimental measurements of magnetic flux evolution with the simulation predictions both without and with the Biermann suppression.
As with CH targets, the simulations without suppression overestimate the flux, though the discrepancy is not as large.
Again, including suppression reduces the spread in the simulation predictions due to tuning the coupling efficiency, from $7\%$ to 2$\%$.
However, for Cu targets the Biermann suppression model reduces the predicted flux below experimental observations.
   
The simulation and experimental results suggest that non-local suppression effects are more significant for low-Z targets.
Without Biermann suppression, simulations with Cu targets predict lower magnetic flux than the CH results.
This is predominantly due to additional radiative losses at higher Z reducing temperature gradients.
In contrast, experimental measurements of the magnetic flux at t$_{0}$+0.75~ns increases from $1\times10^4$~MG$\mu$m$^2$ for CH to $2\times10^4$~MG$\mu$m$^2$ for Cu.
The same qualitative trend is also seen in the simulations including Biermann suppression. 
The copper targets are less kinetic, due to both lower temperature gradients from radiative losses and shorter mean-free paths for higher Z plasmas.

\begin{figure}
    \centering
    \includegraphics[]{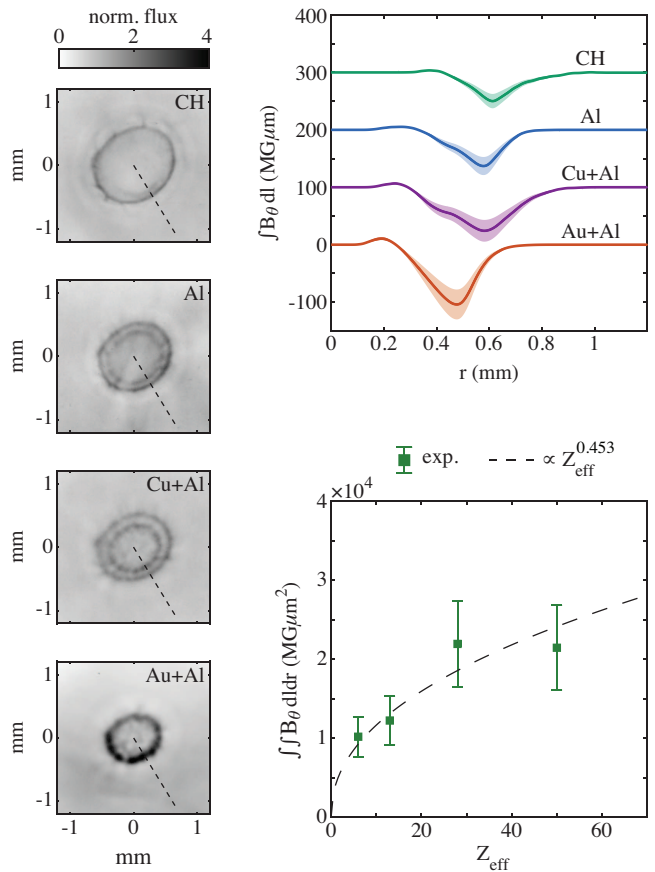}
    \caption{Comparison of experimental results for CH, Al, Cu+Al, and Au+Al targets at t$_0$+0.75~ns.
    The left column shows proton images, and line-out locations are indicated by dashed lines.
    Reconstructed magnetic field profiles are plotted in the top right panel.
    Magnetic flux measurements for each material are plotted as a function of Z$_{\textrm{eff}}$ in the bottom right panel.
    Note, for Cu the value of Z$_{\textrm{eff}}$ = 28, and Z$_{\textrm{eff}}$ is set to 50 for Au.
    The dashed line shows a scaling for flux generation as a function of Z$_{\textrm{eff}}$ based on equation \ref{eq:sherlock_sup}.}
    \label{fig:Z_flux}
\end{figure}

The influence of target material on the magnetic field structure and flux is illustrated in more detail in Figure~\ref{fig:Z_flux}.
The left column shows experimental proton deflectometry images for CH, Al, Cu+Al, and Au+Al targets at t$_0$+0.75~ns.
The proton energies are 37.3~MeV for CH and Al, 32.8~MeV for Cu+Al, and 30.7~MeV for Au+Al.
The reconstructed magnetic field profiles plotted in the top panel of the right column show the Z-dependence of field profiles.
By t$_0$+0.75~ns, the fields from a CH target have expanded the furthest, while the Au+Al has the narrowest features --- indicative of slower plasma evolution due heavier, higher Z ions.
The mid-Z targets both show evidence of magnetic signatures of DAF structures (visible both in the deflectometry and the reconstructed magnetic field profiles) \cite{Campbell_PRL_2020}.

The measured magnetic flux is plotted as a function of Z$_{\textrm{eff}}$ in the lower right panel of Figure \ref{fig:Z_flux}.
While CH and Al are expected to be fully ionized, Cu+Al is assumed to be H-like (Z$_{\textrm{eff}} = 28$)\cite{Campbell_PRL_2020}, and Z$_{\textrm{eff}} = 50$ is used for Au+Al, consistent with Refs.\ \onlinecite{Sherlock_PRL_2020} and \onlinecite{Jones_POP_2017_Au_modeling}.
The measured flux increases with Z$_{\textrm{eff}}$ moving from low-Z to mid-Z targets before slightly decreasing or plateauing for the highest Z.
The observed decrease at high-Z could be a result of slower plasma evolution (also evident from the narrow radius discussed above), or due to more of the coupled laser energy driving ionization and radiation emission, reducing the peak electron temperature.

An estimate of a Z$_{\textrm{eff}}$ scaling taking into account Biermann suppression is plotted with a dashed line.
Based on equation \ref{eq:sherlock_sup} assuming a fixed temperature profile, shorter mean-free-paths lead to less suppression of classical Biermann generation rates ($\lambda_{ei} \propto 1/Z$).
While the scaling appears provide a reasonable description of the data, the assumption of fixed temperature gradients is overly simple considering experimental evidence of temperature gradient changes across the different materials (DAF structures in mid-Z, slower expansion in high-Z).
In addition, the implementation of equation \ref{eq:sherlock_sup} in Gorgon shown in Figures \ref{fig:CH_flux} and \ref{fig:Cu_flux} overestimates the flux for CH, while underestimating for Cu. 

Overall, this suggests additional physics is contributing to the energy transport, and field generation is not accurately modeled by Gorgon, or captured by the non-local Biermann suppression approximated by equation \ref{eq:sherlock_sup}.
In particular, the details of the radiation-hydrodynamics and equation-of-state likely influence plasma dynamics.
Additionally, deviations from a Maxwellian distribution driven by inverse-bremsstrahlung heating can impact energy transport \cite{Ridgers_POP_2008}.
Exploration of such effects is beyond the scope of this work as additional experimental work is needed to carefully constrain the plasma conditions, especially the temperature and density profiles.
Still, from the magnetic field measurements presented here, it is evident that fields are suppressed below classical predictions and that the suppression effect is stronger for lower-Z plasmas.

\section{Conclusion}

Quantitative measurements of magnetic flux enable detailed comparisons between experiments and extended-MHD simulations, demonstrating the need to account for suppression of Biermann battery generation due to non-local effects.
Even with the Biermann suppression, the simulations with CH targets still predict larger magnetic flux than observed experimentally.
However for Cu, while some suppression is necessary, the implementation of equation \ref{eq:sherlock_sup} decreases the flux below experimental observations.
Nevertheless, experimental measurements of magnetic flux as a function of Z$_{\textrm{eff}}$ shows reasonable agreement with the mean-free-path scaling predicted by Ref.\ \onlinecite{Sherlock_PRL_2020}.
The effects of radiation-hydrodynamics and the equation-of-state likely influence the details of simulations, but are beyond the scope of the work.
In future experiments, additional diagnostics, such as Thomson scattering and interferometry, can help constrain plasma parameters to further validate and improve extended-MHD models.
Together with the magnetic field analysis presented in this work, measurements of the temperature and density profiles can elucidate the dynamic interplay between energy transport and field generation in HED plasmas.

\acknowledgments{This material is based upon work supported by the Department of Energy, National Nuclear Security Administration under Award Numbers DE-NA0003606, DE-NA0003764 and DE-AC52-07NA27344. 
PTC is supported by the U.S. Department of Energy Fusion Energy Sciences Postdoctoral Research Program administered by the Oak Ridge Institute for Science and Education (ORISE) for the DOE. ORISE is managed by Oak Ridge Associated Universities (ORAU) under DOE contract number DE-SC0014664. All opinions expressed in this paper are the author's and do not necessarily reflect the policies and views of DOE, ORAU, or ORISE.
BKR acknowledges support from NSF Award Number 1751462. 

This document was prepared as an account of work sponsored by an agency of the United States government. Neither the United States government nor Lawrence Livermore National Security, LLC, nor any of their employees makes any warranty, expressed or implied, or assumes any legal liability or responsibility for the accuracy, completeness, or usefulness of any information, apparatus, product, or process disclosed, or represents that its use would not infringe privately owned rights. Reference herein to any specific commercial product, process, or service by trade name, trademark, manufacturer, or otherwise does not necessarily constitute or imply its endorsement, recommendation, or favoring by the United States government or Lawrence Livermore National Security, LLC. The views and opinions of authors expressed herein do not necessarily state or reflect those of the United States government or Lawrence Livermore National Security, LLC, and shall not be used for advertising or product endorsement purposes.

}

\section*{Data Availability}

The data that support the findings of this study are available from the corresponding author upon reasonable request.

\bibliography{compressed_bib}

\end{document}